\documentclass[12pt]{article}
\usepackage{amssymb,amsmath,epsfig}
\usepackage{cite}
\allowdisplaybreaks

\begin{document}
\title{\bf Role of $f(R,T,R_{\mu\nu}T^{\mu\nu})$ Model on the Stability of Cylindrical Stellar Model}

\author{Z. Yousaf \thanks{zeeshan.math@pu.edu.pk}, M. Zaeem-ul-Haq Bhatti
\thanks{mzaeem.math@pu.edu.pk} and Ume Farwa \thanks{ume.farwa514@gmail.com}\\
Department of Mathematics, University of the Punjab,\\
Quaid-i-Azam Campus, Lahore-54590, Pakistan.}

\date{}

\maketitle
\begin{abstract}
The aim of this paper is to investigate the stable/unstable regimes
of the non-static anisotropic filamentary stellar models in the
framework of $f(R,T,R_{\mu\nu}T^{\mu\nu})$ gravity. We construct the
field equations and conservation laws in the perspective of this
gravity. The perturbation scheme is applied to analyze the behavior
of a particular $f(R,T,R_{\mu\nu}T^{\mu\nu})$ cosmological model on
the evolution of cylindrical system. The role of adiabatic index is
also checked in the formulations of instability regions. We have
explored the instability constraints at Newtonian and post-Newtonian
limits. Our results reinforce the significance of adiabatic index
and dark source terms in the stability analysis of celestial objects
in modified gravity.
\end{abstract}
{\bf Keywords:} $f(R,T,Q)$ gravity; Non-dissipative fluid; Adiabatic index.\\
{\bf PACS:} 04.20.-q; 04.40.-b; 04.40.Dg; 04.40.Nr.

\section{Introduction}

The accelerated expansion of the cosmos is strongly manifested after
the discovery of unexpected reduction in the detected energy fluxes
coming from Supernovae type Ia \cite{snIa,snIb}. Other observational
data like cosmic microwave background radiations, large scale
structures and galaxy red shift surveys \cite{cmb1,cmb2} also
provide evidence in this favor. These observational data propose an
enigmatic form of force, dubbed as dark energy (DE) which takes part
in the expansion phenomenon and dominates overall energy density of
the cosmos. Despite some very solid
claims about the existence of DE, its unknown nature is the
substantial puzzle in cosmology. The idea of modified gravitational
theories is obtained by extending the standard Einstein-Hilbert (EH)
action, which has gained much fame in order to demonstrate the
secrets of cosmic accelerating expansion.

There exists various theories of modified gravity such as $f(R)$
gravity with $R$ as the Ricci scalar, $f(\mathcal{T})$ gravity in
which $\mathcal{T}$ is a torsion scalar, $f(R,T)$ gravity with $T$
as the trace of energy-momentum tensor, $f(G)$ gravity in which $G$
represents Gauss-Bonnet invariant and $f(R,T,Q)$ gravity (where
$Q=R_{\lambda\sigma} T^{\lambda\sigma}$) etc. Nojiri and Odintsov
\cite{z5a} reviewed various versions of modified gravity models that
could explain DE dominance in this accelerating cosmos. Cognola
\emph{et al.} \cite{z5b} introduced some viable formulations of
$f(R)$ DE models and classified them into four main-streams. Nojiri
and Odintsov \cite{b2b} studied some important aspects of $f(R)$
gravity in order to make them well-consistent with observational
data. Bamba \emph{et al.} \cite{z5d} discussed the role of DE,
through modified cosmic models, on the expansion of our accelerating
cosmos. Durrer and Maartens \cite{6} investigated that some $f(R)$
models could lead to new schemes to test out the credibility of
general relativity itself on cosmological scales. Bhatti \emph{et
al.} \cite{7} discussed dynamical instability of non-static
cylindrical cosmic configuration by using $f(\mathcal{T})$ gravity
and found that additional curvature conditions generate the
stability of expanding stellar frame.

Harko \emph{et al.} \cite{8} used $f(R,T)$ theory of gravity and
presented the corresponding equations of motion for the massive
particles through variational principle in $f(R,T)$ theory. The
generalization of $f(R,T)$ gravity is $f(R,T,Q)$ gravity, where
$Q=R_{\lambda\sigma} T^{\lambda\sigma}$ shows the non-minimal
coupling between matter and geometry \cite{8a}. Haghani \emph{et
al.} \cite{19} obtained the field equations by using Lagrange
multiplier in $f(R,T,Q)$ theory of gravity. Odintsov and
S\'{a}ez-G\'{o}mez \cite{20} studied $f(R,T,Q)$ gravity with
non-minimal association between matter and gravitational fields and
concluded that respective modified gravity contains additional
points which would recast the possible cosmological evolution.
Elizalde and Vacaru \cite{z20a} evaluated some exact off-diagonal
cosmological models in $f(R,T,Q)$ gravity. Baffou \emph{et al.}
\cite{z20b} used perturbation technique and performed stability
analysis with the help of de-Sitter and power law models through
numerical simulations in $f(R,T,Q)$ gravity.

Gravitational collapse is the fundamental and highly dissipative
phenomena for structure formation in our universe. Chandrasekhar
\cite{17} studied the dynamical instability of oscillating
spherically symmetric model by using perfect fluid and found
instability limits in terms of adiabatic index. Herrera \emph{et
al.} \cite{18} analyzed the dynamical instability of dense stars
with zero expansion scalar in spherically symmetric configuration
and found instability limits that are independent of adiabatic
index. Cembranos \emph{et al.} \cite{14} studied gravitational
collapse in $f(R)$ gravity and found this phenomenon as a key tool
to constrain modified gravity models that depict late time
cosmological acceleration. Yousaf \emph{et al.} \cite{10}
investigated the irregularity constituents for spherical
self-gravitating stars in the presence of imperfect matter
distribution within $f(R,T)$ gravity and found that the complexity
of matter increases with the increase of anisotropic stresses.
Yousaf \cite{zs1} explored collapsing spherical models supporting
vacuum core in a $\Lambda$-dominated era within the stellar
interior.

The subject of exploring the cosmic filamentary celestial objects
has been a source of great attention, motivated by many relativistic
astrophysicists \cite{bin4,bin6,bin7,bin5,bin3,bin7a}. On large
cosmic scale, it has been analyzed that matter is usually configured
to make large filaments. These stellar structures have been found to
be very clear characteristics of the interstellar medium. They may
give rise to galaxies upon contraction. Motivated from several
simulations and observational outcomes, the stability analysis of
cosmic filaments with more realistic assumptions has received great
interest. Binney and Tremaine \cite{bin1} have linearized the Vlasov
equation about the steady phase of the relativistic interior and
solved the resulting eigenvalue equation in order to discuss the
dynamical stability of collision-less celestial structures supported
by the Vlasov-Poisson formulations. Chavanis \cite{bin2} has
extended their results in the context of non-linear dynamical
stability and explored the problem of stability of barotropic as
well as collision less stellar systems via the maximization of a
Casimir functional (or H-function) with fixed values of energy and
mass. Quillen and Comparetta \cite{bin8} assumed a constant linear
mass density and approximately evaluated a dispersion relation in
the background of the tidal galaxy tail.

Myers \cite{bin10} has discussed the evolution of some observed
characteristics of cores and filaments and concluded that during the
contraction of host filaments, the core grows in mass and radius and
this phenomenon stops if the surrounding filament gas will no longer
exists for further accretion. Breysse \emph{et al.} \cite{bin11}
carried out analytical approach with the detailed perturbation
background and investigated the stability of polytropic fluid
filaments. They found that the instabilities of the cosmic fluid
filaments could be enhanced by introducing tangential fluid motion
of the system. Sharif and Manzoor \cite{chi4} studied the dynamical
instability of axially symmetric stellar structure with reflection
degrees of freedom coupled with locally anisotropic fluid
configurations in self-interacting Brans-Dicke gravity and obtained
stability conditions through adiabatic index at both N and pN
approximations. Birnboim \emph{et al.} \cite{bin9} performed
stability analysis in planar, filamentary and spherical infall
geometries for the existence of virialized gas in one, two and
three-dimensional (3D) gravitational collapse and concluded that
cosmic filaments are likely to host halos under some constraints.

Recently, we have investigated the anisotropic spherical collapse in
the background of $f(R,T,Q)$ gravity and discussed the stability of
compact stars by taking into account the particular viable model
with perturbation technique. We also examined that adiabatic index
$\Gamma_1$ has significant role in the dynamical instability of
these massive stars \cite{22}. The motivation of this paper is to
explain the mathematical as well as physical features of self
gravitating cylindrical celestial objects within the framework of
$f(R,T,Q)$ theory of gravity. Particularly, some properties of
viable modified gravity model are discussed to create the expansion
and DE consequences in cosmos. This paper is organized as follows:
We provide the basic formalism of $f(R,T,Q)$ gravity in section
\textbf{2}. Section \textbf{3} deals with the dynamics of
cylindrical self-gravitating collapsing model in which formation of
field equation and conservation laws by linear perturbation
technique and instability constraints at Newtonian (N) and
post-Newtonian (pN) limits are investigated. Finally, we conclude
our main results in the last section.

\section{The Formalism of $f(R,T,Q)$ Gravity}

The formalism of $f(R,T,Q)$ gravity is based on the contribution of
non-minimal coupling of geometry and matter. Where the $R$ in EH
action is replaced with an arbitrary function of $R,~ T$ and
$R_{\gamma\delta}T^{\gamma\delta}$. In \cite{19} modified EH action
is demonstrated in the following way
\begin{equation}\label{1}
I_{f(R,T,Q)}=\frac{1} {2}\int d^4x\sqrt{-g}
[f(R,T,R_{\lambda\sigma}T^{\lambda\sigma})+ \textit{L}_m] ,
\end{equation}
where, $\textit{L}_m$ expresses the relative Lagrangian density of
matter distribution then the respective energy momentum tensor is
expressed as
\begin{align}\label{2}
&T_{\lambda\sigma}^{(m)}=-\frac{2}{\sqrt{-g}}\frac{\delta(\sqrt{-g}\textit{L}_m)}
{\delta{g^{\lambda\sigma}}}.
\end{align}
On varying the modified action  Eq.(\ref{1}), with metric tensor
$g_{\lambda\sigma}$, the following field equations are obtained
\begin{align}\nonumber
&-G_{\lambda\sigma}(f_{Q}\mathcal{L}_m - f_R) - g_{\lambda\sigma}
\left\{\frac{f} {2}-\Box f_R
-\frac{R}{2}f_R-\frac{1}{2}\nabla_\pi\nabla_\rho(f_{Q}T^{\pi\rho})
-\mathcal{L}_mf_T\right\}\\\nonumber
&+2f_QR_{\pi(\lambda}T_{\sigma)}^{~\pi}+\frac{1}{2}\Box(f_QT_{\lambda\sigma})-\nabla_\pi\nabla_{(\lambda}
[T^\pi_{~\sigma)}f_Q]-2\left(f_Tg^{\pi\rho}+f_QR^{\pi\rho}\right)\frac{\partial^2\mathcal{L}_m}{\partial
g^{\lambda\sigma}\partial g^{\pi\rho}}\\\label{3} &-
T_{\lambda\sigma}^{(m)}(f_T+\frac{R}{2}f_Q+1)-\nabla_\lambda\nabla_\sigma
f_R=0,
\end{align}
where $\nabla_\pi$ and $G_{\lambda\sigma}$ indicates covariant
derivative and Einstein tensor, respectively, with
$\Box=g^{\lambda\sigma}\nabla_\lambda\nabla_\sigma$ as a
d'Alembert's operator. From  Eq.(\ref{3}), one can obtain the
expression of trace as in \cite{22}. In the framework of \cite{8}
the matter Lagrangian has no specific distinction for perfect fluid,
and the corresponding second variation was neglected in their
calculations. Equation (\ref{3}) can be rewritten in GR perspective
as follows
\begin{align}\label{4}
&R_{\lambda\sigma}-\frac{R}{2}
g_{\lambda\sigma}=G_{\lambda\sigma}=\overset{~~\textrm{eff}}{{T}_{\lambda\sigma}},
\end{align}
where the effective energy momentum tensor
${{T}_{\lambda\sigma}}^{\textrm{eff}}$ has the following form
\begin{align}\nonumber
{{T}_{\lambda\sigma}}^{\textrm{eff}}&=\frac{1}{(f_R-f_Q\textit{L}_m)}\left
[(f_T+\frac{1}{2}Rf_Q+1)T^{(m)}_{\lambda\sigma}+
\left\{\frac{R}{2}\left(\frac{f}{R}-f_R\right)-\textit{L}_mf_T-\frac{1}{2}\right.\right.\\\nonumber
&\left.\left.\times\nabla_{\pi}\nabla_{\rho}(f_QT^{\pi\rho})\right\}g_{\lambda\sigma}
-\frac{1}{2}\Box(f_QT_{\lambda\sigma})
-(g_{\lambda\sigma}\Box-\nabla_\lambda\nabla_\sigma)f_R-2f_QR_{\pi(\lambda}T^\pi_{~\sigma)}\right.\\\nonumber
&\left.+\nabla_\pi\nabla_{(\lambda}
[T^\pi_{~\sigma)}f_Q]+2\left(f_QR^{\pi\rho}+f_Tg^{\pi\rho}\right)\frac{\partial^2\textit{L}_m}{\partial
g^{\lambda\sigma}\partial g^{\pi\rho}}\right].
\end{align}
On taking $Q=0$ in above equation, then $f(R,T,Q)$ gravity would
reduce to $f(R,T)$ theory. However, in case of vacuum it leads to
$f(R)$ gravity theory and consequently we will obtain GR results
whenever $f(R)=R$.

\section{Anisotropic Matter Distribution and Cylindrical Field Equations}

We consider the 3-dimensional (3D) timelike hypersurface, $\Delta$,
that would demarcate the 4D manifold $\mathcal{W}$ into couple of
regions, i.e., exterior $\mathcal{W}^+$ and interior
$\mathcal{W}^-$. The interior region of relativistic stellar system
is given by the following cylindrically symmetric spacetime
\begin{equation}\label{5}
ds^2_-=-A^2(t,r)(dt^{2}-dr^{2})+B^2(t,r)dz^{2}+C^2(t,r)d\phi^{2}.
\end{equation}
For the representation of cylindrical symmetry, the following ranges
are imposed on the coordinates
$$-\infty\leq t\leq\infty, ~~ 0\leq r, ~~ -\infty<z<\infty, ~~
0\leq\phi\leq2\pi.$$ We number the respective coordinates $x^0=t$,
$x^1=r$, $x^2=z$ and $x^3=\phi$. We assumed $C = 0$ at $r = 0$ that
represents a non-singular axis. The spacetime for $\mathcal{W}^+$ is
\cite{v34}
\begin{equation}\label{6n}
ds^2_+=-e^{2(\gamma-\upsilon)}(d\nu^2-d\rho^2)+e^{-2\upsilon}\rho^2d\phi^2+e^{2\upsilon}d{z}^2,
\end{equation}
where $\gamma$ and $\upsilon$ are the functions of $\nu$ and $\rho$,
while the coordinates are numbered as
$x^{\beta}=(\nu,~\rho,~\phi,~z)$. The corresponding vacuum field
equations provide
\begin{align}\label{k1}
&\rho(\upsilon_\nu^2+\upsilon_\rho^2)=\frac{\tilde{f}-\tilde{R}\tilde{f}_R}{2\tilde{f}_R}e^{2(\gamma-\upsilon)},\\\label{k2}
&2\rho\upsilon_\nu\upsilon_\rho=\gamma_\nu,\\\label{k3}
&\upsilon_{\nu\nu}-\frac{\upsilon_{\rho}}{\rho}-\upsilon_{\rho\rho}=\frac{e^{2(\gamma-\upsilon)}}{4\rho}
\left(\frac{\tilde{f}-\tilde{R}\tilde{f}_R}{\tilde{f}_R}\right)\left\{\rho
e^{-4\upsilon}+\frac{e^{2\gamma}}{\rho}\right\},
\end{align}
where subscripts $\rho$ and $\nu$ show partial differentiations with
respect to $\rho$ and $\nu$, respectively and tilde indicates that the corresponding values are evaluated
with constant $R,~T$ and $Q$ conditions. It has been proved by Senovilla \cite{sano1} that modified extra curvature terms on the boundary surface should be constant.
Due to this reason, we have evaluated above equations with constant $R,~T$ and $Q$.
These equations suggest the existence of gravitational field. We assume anisotropic and
non-dissipative collapsing matter in cylindrical geometry, whose
energy momentum tensor is
\begin{equation}\label{6}
T_{\lambda\sigma}=(P_{r}+\mu)V_{\lambda}V_{\sigma}+P_{r}g_{\lambda\sigma}
-K_\lambda
K_\sigma(P_{r}-P_{\phi})-S_{\lambda}S_{\sigma}(P_{r}-P_{z}),
\end{equation}
where $\mu$ is the energy density which is the eigenvalue of
$T_{\lambda\sigma}$ for eigenvector $V_\lambda$, while
$P_{\phi},~P_z,~P_r$ are the principal stresses. The spacetime
(\ref{5}) is the canonical form for cylindrical symmetry, defined as
usual by the 2D group that defines the cylindrical symmetry. The
unitary vectors $V_\lambda,~L_\lambda,~S_\lambda,~K_\lambda$ are
configuring to make a canonical orthonormal tetrad in which a
hypersurface orthogonal 4-velocity vector is $V_\lambda$. Further,
the two vectors $S_\lambda$ and $K_\lambda$ are tangent to the
orbits of the 2D group that preserves cylindrical geometry and
$L_\lambda$ is orthogonal to 4-velocity $V_\lambda$ and to these
orbits. It is worthy to stress that we are considering an Eckart
frame where fluid elements are at the state of rest. The four
vectors obey the following relations
\begin{equation}\label{7}
V^{\lambda}V_{\lambda}=-1, ~
K^{\lambda}K_{\lambda}=1=S^{\lambda}S_{\lambda}, ~~
V^{\lambda}K_{\lambda}=V^{\lambda}S_{\lambda}=K^{\lambda}S_{\lambda}=0.
\end{equation}
We choose the fluid to be comoving in a given coordinate system,
therefore, we have
\begin{equation}\label{8}
V_{\lambda}=-A\delta^{0}_{\lambda},~~
K_{\lambda}=C\delta^{3}_{\lambda},~~
L_{\lambda}=A\delta^{1}_{\lambda}~~\textrm{and}~~
S_{\lambda}=B\delta^{2}_{\lambda},
\end{equation}
The four acceleration vector is
$a_\lambda=V_{\lambda;\sigma}V^\sigma$, with $a=\frac{A'}{A^2}$ as a
scalar associated with the four-acceleration. The expansion scalar,
($\Theta=V^{\lambda}_{~~;\lambda}$), for our cylindrical spacetime
leads to
\begin{equation}\label{9}
\Theta=\frac{1}{A}\left(\frac{\dot{A}}{A}+\frac{\dot{B}}{B}+\frac{\dot{C}}{C}\right),
\end{equation}
where over dot represents the time derivative. The shear tensor
$\sigma_{\lambda\sigma}$ is
\begin{align*}
&\sigma_{\lambda\sigma}=V_{(\lambda;\sigma)}+a_{(\lambda}V_{\sigma)}
-\frac{1}{3}\Theta h_{\lambda\sigma},
\end{align*}
where $h_{\lambda\sigma}$ is a projection tensor with
$h_{\lambda\sigma}=g_{\lambda\sigma}+V_{\lambda}V_{\sigma}$. The
shear tensor can also be expressed as follows
\begin{equation}\nonumber
\sigma_{\lambda\sigma}=\sigma_s\left(S_\lambda
S_\sigma-\frac{h_{\lambda\sigma}}{3}\right)+\sigma_k\left(K_\lambda
K_\sigma-\frac{h_{\lambda\sigma}}{3}\right),
\end{equation}
where
\begin{equation}\label{10}
\sigma_s=-\frac{1}{A}\left(\frac{\dot{A}}{A}-\frac{\dot{B}}{B}\right),
~
\sigma_k=-\frac{1}{A}\left(\frac{\dot{A}}{A}-\frac{\dot{C}}{C}\right).
\end{equation}
The non zero modified gravitational field equations for our
cylindrical line element associated with matter distribution
(\ref{6}) take the form
\begin{align}\label{11}
&\frac{1}{A^2}\left[\frac{\dot{C}\dot{B}}{BC}-\frac{C''}{C}-\frac{B''}
{B}-\frac{B'C'}{BC}+\alpha_1\right]=\overset{\textrm{eff}}{\mu},\\\label{12}
&\left(\frac{B'}{B}+\frac{C'}{C}\right)\frac{\dot{A}}{A}-\frac{
\dot{C'}}{C}-\frac{\dot{B'}}{B}+\left(\frac{\dot{B}}{B}+\frac{\dot{C}}{C}\right)
\frac{A'}{A}=0,\\\label{13}
&\frac{B'{C'}}{BC}-\frac{\ddot{B}}{B}-\frac{\dot{B}\dot{C}}{BC}
-\frac{\ddot{C}}{C}+\alpha_1=\overset{\textrm{eff}}{P_r},\\\label{14}
&\left(\frac{B}{A}\right)^2\left[\beta_1
+\frac{C''}{C}-\frac{\ddot{C}}{C}\right]=\overset{\textrm{eff}}{P_z},\quad
\left(\frac{C}{A}\right)^2\left[\beta_1+\frac{B''}{B}-\frac{\ddot{B}}{B}
\right]=\overset{\textrm{eff}}{P_\phi},\\\nonumber
\end{align}
where
\begin{align}\nonumber
\alpha_1&=\left(\frac{\dot{C}}
{C}+\frac{\dot{B}}{B}\right)\frac{\dot{A}}{A}+\left(\frac{{B'}}{B}+\frac{{C'}}{C}\right)\frac{{A'}}{A}
,~\beta_1=\frac{\dot{A}^2}{A^2}-\frac{A'^2}{A^2}-\frac{\ddot{A}}{A}
+\frac{A''}{A},\\\label{11n}
\overset{\textrm{eff}}{\mu}&=\frac{1}{f_R-f_Q\mathcal{L}_M}
\left[\mathcal{L}_Mf_T-\frac{1}{2}(f-Rf_R)+\mu\chi_1+\dot{\mu}\chi_2
+\frac{\ddot{\mu}}{2A^2}f_Q+\frac{\mu''}{2A^2}f_Q\right.\\\nonumber
&\left.+\mu'\chi_3+\frac{P''_r}{2A^2}f_Q+P_r\chi_4+P'_r\left\{\frac{f_Q'}{A^2}-\frac{5A'}{2A^3}f_Q\right\}
-\frac{f_Q}{2A^2B}(\dot{P_z}\dot{B}+P'_zB')\right.\\\nonumber
&\left.-\dot{P_r}\frac{\dot{A}}{A^3}f_Q+P_z\chi_5+P_\phi\chi_6-\frac{f_Q}{2A^2}\left(\dot{P_\phi}\frac{\dot{C}}{C}
-{P_\phi'}\frac{C'}{C}\right)-\frac{\dot{f_R}}{A^2}\left(\frac{\dot{A}}{A}+
\frac{\dot{B}}{B}+\frac{\dot{C}}{C}\right)\right.\\\nonumber
&\left.-\frac{{f_R'}}{A^2}\left(\frac{{A'}}{A}-
\frac{{B'}}{B}-\frac{{C'}}{C}\right)+\frac{f_R''}{A^2}\right],\\\nonumber
\overset{\textrm{eff}}{P_r}&=\frac{1}{f_R-f_Q\mathcal{L}_M}\left[\frac{1}{2}(f-Rf_R)
-\mathcal{L}_Mf_T+\frac{\ddot{f_R}}{A^2}\psi_1+\dot{\mu}
\left(\frac{5\dot{A}}{2A^3}f_Q-\frac{\dot{f_Q}}{A^2}\right)\right.\\\nonumber
&\left.-f_R'\psi_2+P_r\chi_7-\frac{\ddot{\mu}}{2A^2}f_Q+\mu\chi_8+\frac{\mu'A'}{2A^3}f_Q
-\frac{\ddot{P_r}}{2A^2}f_Q+P_z\chi_9+P_\phi\chi_{10}\right.\\\nonumber
&\left.+\frac{f_Q}{2A^2}\left\{\dot{P_z}\frac{\dot{B}}{B}-P_z'\frac{B'}{B}
+\dot{P_\phi}\frac{\dot{C}}{C}-P_\phi'\frac{C'}{C}\right\}+\dot{P_r}\chi_{11}
+P_r'\chi_{12}\right],\\\label{12n}
\overset{\textrm{eff}}{P_z}&=\frac{1}{f_R-f_Q\mathcal{L}_M}\left[\frac{1}{2}(f-Rf_R)-\mathcal{L}_Mf_T+\dot{\mu}
\chi_{14}+\mu\chi_{13}-\frac{\ddot{\mu}}{2A^2}f_Q+\frac{\mu'A'}{2A^3}f_Q\right.\\\nonumber
&\left.+P_r\chi_{15}+\frac{\dot{A}\dot{P_r}}{2A^3}f_Q+P_r'\left(\frac{5A'}{2A^3}f_Q-\frac{f'_Q}{A^2}\right)
-\frac{P''_r}{2A^2}f_Q+P_z\chi_{16}+P_z'\chi_{17}-\dot{P_z}\right.\\\label{13n}
&\left.\times\chi_{18}-\frac{\ddot{P_z}}{2A^2}f_Q
+\frac{P''_z}{2A^2}f_Q+P_{\phi}\chi_{19}+\frac{f_Q}{2A^2}\left(\dot{P_{\phi}}\frac{\dot{C}}{C}
-P_{\phi}'\frac{C'}{C}\right)+\psi_3\right],\\\nonumber
\overset{\textrm{eff}}{P_\phi}&=\frac{1}{f_R-f_Q\mathcal{L}_M}\left[\frac{1}{2}(f-Rf_R)-\mathcal{L}_Mf_T+\dot{\mu}
\chi_{14}+\mu\chi_{13}-\frac{\ddot{\mu}}{2A^2}f_Q+\frac{\mu'A'}{2A^3}f_Q\right.\\\nonumber
&\left.+P_r\chi_{15}+\frac{\dot{A}\dot{P_r}}{2A^3}f_Q+P_r'\left(\frac{5A'}{2A^3}f_Q-\frac{f'_Q}{A^2}\right)
-\frac{P''_r}{2A^2}f_Q+P_z\chi_{20}+P_\phi'\chi_{23}+\dot{P_\phi}\right.\\\label{14n}
&\left.\times\chi_{22}-\frac{\ddot{P_\phi}}{2A^2}f_Q
+\frac{P''_\phi}{2A^2}f_Q+P_{\phi}\chi_{21}+\frac{f_Q}{2A^2}\left(\dot{P_{z}}\frac{\dot{B}}{B}
-P_{z}'\frac{B'}{B}\right)+\psi_4\right],
\end{align}
where prime stands for $\frac{\partial}{\partial r}$ operator and
the quantities $\chi_i$'s contain combinations of metric variables
and their derivatives are mentioned in Appendix. The value of $R$
for respective spacetime is given as
\begin{align}\nonumber
R&=\frac{2}{A^2}\left[\left(\frac{\ddot{A}}{A}+\frac{\ddot{B}}{B}+\frac{\ddot{C}}{C}\right)
-\left(\frac{A''}{A}+\frac{B''}{B}+\frac{C''}{C}\right)
+\frac{1}{A^2}(A'^2-\dot{A}^2)\right.\\\label{15}
&\left.+\frac{1}{BC}(\dot{B}\dot{C} -B'C')\right].
\end{align}

\subsection{Viability Of $f(R,T,Q)$ Model and Junction Conditions}

In this subsection, we shall deal with the hydrodynamics of
cylindrical stellar collapse by using dynamical equations. The
expression of covariant derivative of effective energy momentum
tensor is
\begin{align}\label{16}
\nabla^\lambda
T_{\lambda\sigma}&=\frac{2}{Rf_Q+2f_T+1}\left[\nabla_\sigma(\mathcal{L}_mf_T)
+\nabla_\sigma(f_QR^{\pi\lambda}T_{\pi\sigma})-\frac{1}{2}(f_Tg_{\pi\rho}+f_QR_{\pi\rho})\right.\\\nonumber
&\times\left.\nabla_\sigma
T^{\pi\rho}-G_{\lambda\sigma}\nabla^\lambda(f_Q\mathcal{L}_m)\right],
\end{align}
which would provide two equations of motion in $f(R,T,Q)$ theory.
Making use of $G^{\lambda\sigma}_{~;\sigma}=0$ and
Eqs.(\ref{11})-(\ref{14}) along with $\lambda=0,~1$, the above
equation gives
\begin{align}\nonumber
&\frac{\overset{~~\textrm{eff}}{\dot{\mu}}}{A}+\Theta\left[\overset{~~\textrm{eff}}{\mu}
+\frac{1}{3}\left(\overset{~~\textrm{eff}}{P_{r}}+\overset{~~\textrm{eff}}{P_{z}}
+\overset{~~\textrm{eff}}{P_{\phi}}\right)\right]+\frac{1}{3}\left(\overset{~~\textrm{eff}}{P_{z}}
-\overset{~~\textrm{eff}}{P_{r}}\right)(2\sigma_s-\sigma_k)\\\label{17}
&+\frac{1}{3}\left(\overset{~~\textrm{eff}}{P_{\phi}}
-\overset{~~\textrm{eff}}{P_{r}}\right)(2\sigma_k-\sigma_s)+Z_1=0,\\\nonumber
&\nabla
\overset{~~\textrm{eff}}{P_{r}}-\frac{1}{A}\left[\left(\overset{~~\textrm{eff}}{P_{z}}
-\overset{~~\textrm{eff}}{P_{r}}\right)\frac{B'}{B}+\left(\overset{~~\textrm{eff}}{P_{\phi}}
-\overset{~~\textrm{eff}}{P_{r}}\right)\frac{C'}{C}\right]+\left(\overset{~~\textrm{eff}}{\mu}
-\overset{~~\textrm{eff}}{P_{r}}\right)\\\label{18} &\times a+Z_2=0,
\end{align}
where superscript $``\textrm{eff} "$ indicates the presence of
$f(R,T,Q)$ terms in the matter variables and the expressions of
$Z_1$ and $Z_2$ are mentioned in Appendix as Eqs.(\ref{z1}) and
(\ref{z2}). The quantities $Z_1$ and $Z_2$ are coming due to the
non-conserved divergence of energy momentum tensor. The dynamical
equations could help to explain hydrodynamics of locally anisotropic
cylindrical relativistic massive bodies. It is worthy to mention
that the theoretically designed stellar models are of worth
importance if they are stable against instabilities and
fluctuations. Now, we will explain the dynamic instability of
anisotropic and non dissipative relativistic cylindrical geometry by
using particular $f(R,T,Q)$ model \cite{23}.
\begin{equation}\label{19}
f(R,T,Q)=\alpha R^2+\beta Q,
\end{equation}
where $\alpha$ and $\beta$ are constants. The model, $\alpha
R^n+\beta Q^m$ is the generalization of above mentioned $f(R,T,Q)$
model in which $m$ and $n$ are constants. In order to deal this theory free from
Ostrogradski instabilities, one should take $n\neq1$. However, this model will generate stable theory for
$m=1$, by giving EH term including canonical scalar field having non-minimal variation
coupling of Einstein tensor. The model with $n=2$ and $m=1$ along with constant $\beta$ could help to understand the dynamics and evolution of inflationary
cosmos. For the particular value of constant $\alpha$, i.e.,
$\alpha=\frac{1}{6M^2}$ \cite{24} with $M=2.7\times 10^{-12}$ GeV,
this model behaves as a substitute of DM. In case of
$\alpha=0$, there is geometry-matter association on behalf of
coupling between stress-energy tensor and the Ricci scalar. Yousaf \emph{et al.} \cite{22} studied this model with $n=2,
~m=1$ and discussed the stability of compact stars in anisotropic
spherical configuration by taking $\beta>0$ along with
$\alpha=\frac{1}{6M^2}$.

For the smooth matching of Eqs.(\ref{5}) and (\ref{6n}) over
$\Delta$, we shall use junction conditions proposed by Darmois
\cite{v36} as well as Senovilla \cite{sano1} for $f(R,T,Q)$ theory. Since we have assumed
a timelike hypersurface, therefore we impose $r=$constant in
Eq.(\ref{5}) and $\rho(\nu)$ in the exterior metric (\ref{6}). In
this framework, the first fundamental form provides
\begin{align}\label{22n}
&d\tau\overset{\Delta}=e^{2\gamma-2\upsilon}\left\{1-\left(\frac{d\rho}{d\nu}\right)^2\right\}^{1/2}d\nu=Adt,\\\label{23}
&B\overset{\Delta}=e^{\upsilon},\quad
C\overset{\Delta}=e^{-\upsilon}\rho,
\end{align}
with $1-\left(\frac{d\rho}{d\nu}\right)^2>0$. Here, the notation
overset $\Delta$ indicates that the corresponding equations and
quantities are evaluated on the hypersurface, $\Delta$. The second
fundamental form yields
\begin{align}\label{24n}
&e^{2\gamma-2\upsilon}[\nu_{\tau\tau}\rho_{\tau}-\rho_{\tau\tau}\nu_{\tau}
-\{\nu_\tau(\gamma_\rho-\upsilon_\rho)+\rho_\tau(\gamma_\nu-\upsilon_\nu)\}(\nu_\tau^2-\rho_\tau^2)]
\overset{\Delta}=\frac{-A'}{A^2},\\\label{25n}
&e^{2\upsilon}(\rho_\tau\upsilon_{\nu}+\nu_\tau\upsilon_\rho)\overset{\Delta}=\frac{BB'}{A},\quad
e^{-2\upsilon}\rho^2\left(\rho_\tau\upsilon_\nu+\nu_\tau\upsilon_\rho-\frac{\nu_\tau}{\rho}\right)\overset{\Delta}=\frac{-CC'}{A}.
\end{align}
By making use of Eqs.(\ref{22n})-(\ref{25n}), field equations and
after some manipulations, we obtain
\begin{align}\label{26n}
&\overset{~~\textrm{eff}}{P_r}\overset{\Delta}=0.
\end{align}
From Eq.(\ref{4}), one can write the following form
\begin{align}\nonumber
G_{\lambda\sigma}&=\frac{1}{(f_R-f_Q\textit{L}_m)}\left
[(f_T+\frac{1}{2}Rf_Q+1)T^{(m)}_{\lambda\sigma}+
\left\{\frac{R}{2}\left(\frac{f}{R}-f_R\right)-\textit{L}_mf_T-\frac{1}{2}\right.\right.\\\nonumber
&\left.\left.\times\nabla_{\pi}\nabla_{\rho}(f_QT^{\pi\rho})\right\}g_{\lambda\sigma}
-\frac{1}{2}\Box(f_QT_{\lambda\sigma})
+\nabla_\lambda\nabla_\sigma f_R+g^{\pi\nu}\nabla_\pi\nabla_{\nu}(T_{\lambda\sigma}f_Q)\right.\\\nonumber
&\left.-g_{\lambda\sigma}\Box f_R+2f_QR T_{\lambda\sigma}\right],
\end{align}
which can be transformed as
\begin{align}\nonumber
\Omega_{\lambda\sigma}&=\frac{1}{(1+f_T+\frac{5}{2}R f_Q)}\left[(f_R-f_QL_m)G_{\lambda\sigma}-\frac{1}{2}(f-Rf_R)g_{\lambda\sigma}+L_mf_Tg_{\lambda\sigma}
+\frac{1}{2}\right.\\\label{jc1}
&\left.\left.\times\nabla_{\pi}\nabla_{\rho}(f_QT^{\pi\rho})\right\}g_{\lambda\sigma}
-\frac{1}{2}\Box(f_QT_{\lambda\sigma})
-\nabla_\lambda\nabla_\sigma f_R+g_{\lambda\sigma}\Box f_R\right],
\end{align}
where $\Omega_{\lambda\sigma}$ indicates tensor associated with bulk matter. In a Gaussian normal coordinates system,
we have
\begin{align}\nonumber
ds^2&=dy^2+\gamma_{ab}dx^adx^b,
\end{align}
in Which the boundary surface us at $y=0$. In this context, the Ricci scalar takes the form
\begin{align}\label{ricci}
R=2\partial_yK-\frac{4}{3}K^2-K^*_{ab}K^{*ab}-\tilde{R},
\end{align}
where $K_{ab}$ is the extrinsic curvature at the hypersurface, tilde shows the constant choice of the Ricci scalar evaluated through induced
spacetime, while
$K^*_{ab}$ and $K$ are the trace-less and trace components of the extrinsic curvature respectively. The value of
the extrinsic curvature can be expressed
through $\gamma_{ab}$ as $K_{ab}=-1/2\times\partial_y\gamma_{ab}$. The Einstein tensor provides
\begin{align}\nonumber
G_{yy}&=-\frac{1}{2}(K_{\lambda\sigma}K^{\lambda\sigma}+\tilde{R}-K^2),\quad G_{y\sigma}=-\nabla_\nu(K^\nu_{~\sigma}-\delta^\nu_{~\sigma}
K),\\\nonumber
G_{\lambda\sigma}&=\partial_y(K_{\lambda\sigma}-K\gamma_{\lambda\sigma})+\frac{1}{2}\gamma_{\lambda\sigma}(K_{\mu\nu}K^{\mu\nu}+K^2)
+\tilde{G}_{\lambda\sigma}-3KK_{\lambda\sigma}+2K^\nu_{~\lambda}K_{\nu\sigma}.
\end{align}
Now, we split Eq.(\ref{jc1}) into two tensorial quantities as
\begin{align}\label{omega1}
\Omega_{\lambda\sigma}=Q_{\lambda\sigma}+L_{\lambda\sigma},
\end{align} where
\begin{align}\label{qt}
Q_{\lambda\sigma}&=(f_R-f_QL_m)G_{\lambda\sigma}+L_mf_Tg_{\lambda\sigma}-\frac{1}{2}(f-Rf_R)g_{\lambda\sigma},\\\label{lt}
L_{\lambda\sigma}&=\frac{1}{2}\nabla_\mu\nabla_\nu(f_QT^{\mu\nu})g_{\lambda\sigma}-\frac{1}{2}\Box(f_QT_{\lambda\sigma})
-\nabla_\lambda\nabla_\sigma f_R+g_{\lambda\sigma}\Box f_R.
\end{align}
The components of Eq.(\ref{qt}) are obtained as follows
\begin{align}\nonumber
Q_{yy}&=G_{yy}f_R-f_QL_mG_{yy}+\frac{1}{2}(Rf_R-f),\\\nonumber
Q_{y\beta}&=f_RG_{y\beta}-f_QL_mG_{y\beta},\\\nonumber
Q_{\alpha\beta}&=f_RG_{\alpha\beta}-f_QL_mG_{\alpha\beta}+L_m\gamma_{\alpha\beta}f_T-\frac{1}{2}(f-Rf_R)\gamma_{\alpha\beta},
\end{align}
while Eq.(\ref{lt}) provides the following relations
\begin{align}\nonumber
L_{yy}&=-K\partial_y f_R+\tilde{\Box} f_R -\frac{1}{2} \tilde{\Box}(f_QT_{\alpha\beta})+\frac{K}{2}\partial_y(f_QT^{\alpha\beta}),\\\nonumber
L_{y\beta}&=-\partial_\beta\partial_y f_R-K^\mu_{~\beta}\partial_\mu f_R-\frac{1}{2}\tilde{\Box}(f_QT_{y\beta}),\\\nonumber
L_{\alpha\beta}&=-\tilde{\nabla}_{\alpha\beta}+K_{\alpha\beta}\partial_y f_R+\frac{1}{2}\gamma_{\alpha\beta}
\left[\tilde{\nabla}_{\mu\nu}(f_QT^{\mu\nu})-K_{\mu\nu}\partial_y(f_QT^{\mu\nu})\right],\\\nonumber
&-\frac{1}{2}\tilde{\Box}(f_QT_{\alpha\beta})+\gamma_{\alpha\beta}[\tilde{\Box}f_R+\partial_{yy}f_R-K\partial_yf_R].
\end{align}
Now, we compute the $ya$ and $yy$ components of Eq.(\ref{omega1}), which after some simplifications
give rise to
\begin{align}\label{jc2}
\partial_y[(K_{\lambda\sigma}-K\gamma_{\lambda\sigma})f_R+\gamma_{\lambda\sigma}f_{QQ}\partial_y Q+\gamma_{\lambda\sigma}f_{RR}
\partial_y R]=0.
\end{align}
Upon integration across the hypersurface, Eq.(\ref{jc2}) yields
\begin{align}\label{jc3}
[(K_{\lambda\sigma}-K\gamma_{\lambda\sigma})f_R+\gamma_{\lambda\sigma}f_{QQ}\partial_y Q+\gamma_{\lambda\sigma}f_{RR}
\partial_y R]|_-^+=0.
\end{align}
The integration of Eq.(\ref{ricci}) gives $R|_-^+=0$, while the trace and traceless components of Eq.(\ref{jc3}) gives rise to
\begin{align}\label{27n1}
f_{,RR}[\partial_y R|_-^+=0,\quad f_{,RR}K^*_{\lambda\sigma}|_-^+=0,\quad f_{,QQ}[\partial_y Q|_-^+=0,\quad K|_-^+=0.
\end{align}
along with
\begin{align}\label{27n2}
R|_-^+=0,\quad Q|_-^+=0,\quad \gamma_{\lambda\sigma}|_-^+=0.
\end{align}
provide the matching conditions for $f(R,T,Q)$ theory of gravity in which
$f_{,RR}\neq0$ and $f_{,QQ}\neq0$ should be satisfied. The details of this approach in $f(R)$ gravity
has been mentioned in \cite{sano1,clif1,clif2}. Equation (\ref{26n}) arises due to Darmois junction conditions that
indicates that effective radial pressure on $\Delta$ is zero. The
obeying of Eqs.(\ref{27n1}) and (\ref{27n2}) over $\Sigma$ is required for the
continuity of $R$ and $Q$ invariants even for matter thin shells.

\subsection{Perturbation Scheme}

In order to discuss the stability of cylindrical celestial objects,
we shall explore the perturbed form of field as well as dynamical
equations in this section. As perturbation deals with small
variations in a physical system resulted by gravitational effects of
other stellar objects. Therefore, in recent few decades, researchers
are very keen to analyze the stability of the cosmic stellar
filaments against oscillatory motion induced by perturbations. Here,
we use the linear perturbation scheme with very small perturbation
parameter $\epsilon$ so that one can neglect its second and higher
powers. Initially, the celestial system is considered to be in
hydrostatic equilibrium, but with the passage of time passes, it is
subjected to the oscillatory motion. All the metric functions and
fluid parameters can be perturbed as \cite{18}
\begin{align}\nonumber
&A(t,r)=A_o(r)+{\epsilon}\omega(t)a(r),\quad
\mu(t,r)=\mu_o(r)+\epsilon{\bar{\mu}}(t,r),\\\nonumber
&B(t,r)=B_o(r)+{\epsilon}\omega(t)b(r),\quad
P_r(t,r)=P_{ro}(r)+\epsilon{\bar{P_r}}(t,r),\\\nonumber
&C(t,r)=C_o(r)+{\epsilon}\omega(t)c(r),\quad P_\phi(t,r)=P_{\phi
o}(r)+\epsilon{\bar{P_\phi}}(t,r),\\\nonumber
&R(t,r)=R_o(r)+{\epsilon}\omega(t)d(r),\quad
P_z(t,r)=P_{zo}(r)+\epsilon{\bar{P_z}}(t,r),\\\label{20}
\end{align}
By using above perturbation technique along with junction conditions
(\ref{26n}), (\ref{27n1}) and (\ref{27n2}), Eq.(\ref{13}) can be executed in terms
of second order partial differential equation as
\begin{align}\label{21n}
\ddot{\omega}-\chi^2\omega\overset{\Delta}=0,
\end{align}
where
\begin{align}\nonumber
\chi^2&=\left[
\left(\frac{B_o'}{B_o}+\frac{C_o'}{C_o}\right)\left(\frac{a}{A_o}\right)'+
\left(\frac{A_o'}{A_o}+\frac{C_o'}{C_o}\right)\left(\frac{b}{B_o}\right)'
+\left(\frac{B_o'}{B_o}+\frac{A_o'}{A_o}\right)\left(\frac{c}{C_o}\right)'\right]\\\nonumber
&\times \left(\frac{b}{B_o}+\frac{c}{C_o}\right)^{-1}.
\end{align}
The most general solution of the above equation is given by
\begin{equation}\label{34bn}
\omega(t)=\mathfrak{c}_1\exp({\chi}t)+\mathfrak{c}_2\exp(-{\chi}t),
\end{equation}
where $\mathfrak{c}_1$ and $\mathfrak{c}_2$ are arbitrary constants.
Equation (\ref{34bn}) indicates two solutions that are independent
to each other. Here, we wish to explore unstable regimes of
collapsing stellar anisotropic system. Due to this reason, we
consider that our stellar filament is in static equilibrium at large
past time, i.e., $\omega(-\infty)=0$, then with the passage of time
it enters into the present state and goes forward in the phase of
gravitational collapse by decreasing its areal radius. Such model
could be achieved only by taking $\mathfrak{c}_1=-1$ along with
$\mathfrak{c}_2=0$. This would describes the monotonically
decreasing configuration of the solution as the time proceed
forward.

The most general solution of Eq.(\ref{21n}) includes oscillating and
non-oscillating functions that correspond to stable and unstable
configurations of stellar anisotropic filament, respectively. The
choice $\mathfrak{c}_1=+1$ is exactly equivalent to the case if one
absorbs the sign in $a,~b,~c$ and $d$. Our aim is to explore
instability regimes of collapsing stellar interiors, therefore we
have to restrict our perturbations $a,~b,~c$ and $d$ on the boundary
surface as a positive definite in order to make $\chi^2>0$. (This
assumption has been taken by number of relativistic astrophysicists
\cite{18, chi1, chi1a, chi2, chi3, chi4, chi4a, chi5} to discuss
unstable limits of collapsing stellar populations). The required
solution associated with Eq.(\ref{21n}) can be achieved by taking
$\mathfrak{c}_1=-1$ and $\mathfrak{c}_2=0$ as
\begin{equation}\label{21}
\omega(t)\overset{\Delta}=-\exp{({\chi}t)}.
\end{equation}
The perturbed configuration of $f(R,T,Q)$ model is
\begin{align}\label{22}
f=[\alpha R_o^2+\beta Q_o]+\epsilon2\alpha\omega (t) d(r)R_o,
\end{align}
where
\begin{align}\nonumber
R_o=-\frac{2}{A_o}\left[\frac{A_o''}{A_o}+\frac{B_o''}{B_o}
+\frac{C_o''}{C_o}-\frac{A_o'^2}{A_o}+\frac{B_o'C_o'}{B_oC_o}\right].
\end{align}
By using above perturbation scheme, the static forms of $f(R,T,Q)$
field equations are
\begin{align}\nonumber
G_{00}^{(S)}&=\frac{1}{2\alpha
R_o+\beta\mu_o}\left[\mu_o\chi_{1o}+\mu_o'\chi_{3o}+P_{ro}\chi_{4o}
+P_{zo}\chi_{5o}+P_{\phi o}\chi_{6o}+\frac{\alpha}{2}
\left(\frac{4R_o''}{A_o^2}\right.\right.\\\label{23}
&\left.\left.+R_o^2-4R_o'\psi_{2o}\right)+\frac{\beta}{2A_o^2}
\left(\mu_o''+P_{ro}''+P_{zo}'\frac{B'_o}{B_o}+P_{\phi
o}'\frac{C'_o}{C_o}-5P_{ro}'\frac{A'_o}{A_o}\right)-\frac{\beta}{2}Q_o\right],\\\nonumber
G_{11}^{(S)}&=\frac{1}{2\alpha
R_o+\beta\mu_o}\left[\mu_o\chi_{8o}+P_{ro}'\chi_{12o}+P_{ro}\chi_{7o}
+P_{zo}\chi_{9o}+P_{\phi o}\chi_{10o}-\frac{\alpha}{2}
\left(R_o^2\right.\right.\\\label{24}
&\left.\left.-4R_o'\psi_{2o}\right)+\frac{\beta}{2A_o^2}
\left(\mu_{o}'\frac{A'_o}{2A_o}-P_{zo}'\frac{B'_o}{B_o}-P_{\phi
o}'\frac{C'_o}{C_o}\right)+\frac{\beta}{2}Q_o\right],\\\nonumber
G_{22}^{(S)}&=\frac{1}{2\alpha
R_o+\beta\mu_o}\left[\frac{1}{2}(\beta Q_o-\alpha
R_o^2)+\mu_o\chi_{13o}+P_{ro}\chi_{15o}+P_{zo}\chi_{16o}
+P_{zo}'\chi_{17o}\right.\\\label{25} &\left.+P_{\phi
o}\chi_{14o}+\psi_{3o}+\frac{\beta}{2A_o^2}
\left(\mu_{o}'\frac{A'_o}{A_o}+5P_{ro}'\frac{A_o'}{A_o}-P_{ro}''+P_{zo}''
-P_{\phi o}'\frac{C'_o}{C_o}\right)\right],
\end{align}
where superscript $(S)$ indicates static form of Einstein tensors.
Their expressions are given in Appendix as
Eqs.(\ref{sm})-(\ref{spp}). However, the perturbed configuration of
these equations are
\begin{align}\nonumber
{\bar{G}_{00}}&=\frac{1}{2\alpha R_o+\beta\mu_o}\left[\omega(\alpha
d R_o+\mu_ox_1+P_{ro}x_4+P_{zo}x_5+P_{\phi
o}x_6+\mu_o'x_3)+\bar{\mu}\chi_{1o}\right.\\\nonumber
&\left.+\dot{\mu}\chi_{2o}+\bar{\mu}'\chi_{3o}+\bar{P_r}\chi_{4o}
+\bar{P_z}\chi_{5o}+\bar{P_\phi}\chi_{6o}+\frac{\beta}{2A_o}\left(
\ddot{\bar{\mu}}+\bar{\mu}''+\bar{P_r}''-5\bar{P_r}'\frac{A_o'}{A_o}\right.\right.\\\nonumber
&\left.\left.
+\bar{P_\phi}'\frac{C_o'}{C_o}+\bar{P_z}'\frac{B_o'}{B_o}\right)
-\frac{\omega\beta}{A_o^2}\left(\frac{a\mu_o''}{A_o}+\frac{aP_{ro}''}{A_o}
+\frac{bP'_{zo}B'_o}{2B_o^2}+\frac{aP'_{zo}B'_o}{A_oB_o}
+\frac{bP'_{zo}}{2B_o}+P_{\phi o}'\right.\right.\\\nonumber
&\left.\left.\frac{c'}{2C_o}+\frac{cP_{\phi
o}C_o'}{2C_oz}+\frac{aP_{\phi o}C_o'}{2A_oC_o}\right)+5\omega
P_{ro}'\frac{\beta}{2}\left(\frac{b}{A_o^3}\right)'+2\alpha
\omega\left(\frac{d''}{A_o^2}-\frac{2R''_o}{A_o^3}+R'_oy_2\right.\right.\\\label{26}
&\left.\left.+d'\psi_{2o}
\right)-2\alpha\dot{\omega}d\psi_{1o}\right]-\frac{2\alpha\omega d
+\beta\bar{\mu}}{2\alpha
R_o+\beta\mu_o}\overset{~~\textrm{eff}}{{\mu}_o},\\\nonumber
{\bar{G}_{11}}&=\frac{1}{2\alpha
R_o+\beta\mu_o}\left[\omega(P_{ro}x_7-\alpha d
R_o+\mu_ox_8+P_{zo}x_9+P_{\phi o}x_{10}+P_{r
o}'x_{12}\right.\\\nonumber &\left.-2\alpha R'_oy_2)+2\alpha
d\frac{\ddot{\omega}}{A_o^2}-\frac{\beta}{2A_o^2}\left(\ddot{\bar{P_r}}+\ddot{\bar{\mu}}
-\bar{\mu}\frac{A_o'}{A_o}+\bar{P_\phi}'\frac{C_o'}{C_o}
+\bar{P_z}'\frac{B_o'}{B_o}\right)+\bar{P_r}\right.\\\nonumber
&\left.\times \chi_{7o}+\bar{\mu}\chi_{8o}+\bar{P_z}\chi_{9o}
+\bar{P_\phi}\chi_{10o}+\dot{\bar{P_r}}\chi_{11o}+\bar{P_r}'\chi_{12o}
+\frac{\omega\beta}{2A_o^2}\left(\frac{\mu_o'a'}{A_o}+2a\right.\right.\\\label{27}
&\left.\left.\times\frac{P_{zo}'B_o'}{A_oB_o}-\frac{3\mu_o'a}{A_o^2}
+\frac{bP_{zo}'B_o'}{B_o^2}-\frac{b'P_{zo}'}{B_o} +\frac{cP_{\phi
o}'C_o'}{C_o^2}\right)\right]-\frac{2\alpha\omega d
+\beta\bar{\mu}}{2\alpha
R_o+\beta\mu_o}\overset{~~\textrm{eff}}{{P}_{zo}},\\\nonumber
{\bar{G}_{22}}&=\frac{1}{2\alpha
R_o+\beta\mu_o}\left[\omega(\mu_ox_{13}-\alpha
R_od+P_{ro}x_{15}+P_{zo}x_{16}+P_{zo}'x_{17}+P_{\phi
o}x_{19}\right.\\\nonumber &\left.+y_3)
+\bar{\mu}\chi_{13o}+\dot{\bar{\mu}}\chi_{14o}+\bar{P_r}
\chi_{15o}+\bar{P_z}\chi_{16o}+\bar{P_z}'\chi_{17o}
-\bar{{P_z}}\chi_{18o}+\bar{P_\phi}\chi_{19o}\right.\\\nonumber
&\left.
+\frac{\beta}{2A_o^2}\left(\bar{\mu}'\frac{A_o'}{A_o}-\ddot{\bar{\mu}}
+\bar{P_r}'\frac{A_o'}{A_o}-\bar{P_r}''-\ddot{\bar{P_z}}+\bar{P_z}''
-\bar{P_\phi}'\frac{C_o'}{C_o}\right)+\frac{\omega\beta}{2A_o^2}
\left\{\frac{\mu_o'a'}{A_o}\right.\right.\\\nonumber
&\left.\left.-3\mu_o'a\frac{A_o'}{A_o^2}
+5P_{ro}'\frac{a'}{A_o}-15P_{ro}'a\frac{A_o'}{A_o^2}
+\frac{aP_{ro}''}{A_o}-\frac{aP_{zo}''}{A_o}+P_{\phi
o}'\frac{C_o'}{C_o}\left(\frac{c}{C_o}+\frac{2}{A_o}\right.\right.\right.\\\label{28}
&\left.\left.\left.-1\right)\right\}\right]-\frac{2\alpha\omega d
+\beta\bar{\mu}}{2\alpha
R_o+\beta\mu_o}\overset{~~\textrm{eff}}{{P}_{zo}},\\\nonumber
{\bar{G}_{33}}&=\frac{1}{2\alpha
R_o+\beta\mu_o}\left[\omega(\mu_ox_{13}-\alpha
R_od+P_{ro}x_{15}+P_{zo}x_{20}+P_{zo}'x_{23}+P_{\phi
o}x_{21}\right.\\\nonumber
&\left.+y_4)-\frac{\beta}{2A_o^2}\left(\frac{\mu_o'a'}{A_o}-\frac{3\mu_o'a}{A_o^2}
+\frac{5a'P_{ro}'}{A_o}-15P_{ro}'\frac{a}{A_o^2}+\bar{P_r}''+2aP_{ro}''+bP_{zo}'\right.\right.\\\nonumber
&\left.\left.\frac{B_o'}{B_o}
+aP_{zo'}\frac{B_o'}{A_oB_o}-\frac{b'P_{zo}'}{B_o}-\frac{2aP_{zo}''}{A_0}\right)
+\bar{\mu}\chi_{13o}+\dot{\bar{\mu}}\chi_{14o}+\bar{P_r}
\chi_{15o}+\bar{P_z}\chi_{20o}\right.\\\label{29}
&\left.+\bar{P_\phi}\chi_{21o}+
\dot{\bar{P_\phi}}\chi_{22o}+\bar{P_\phi}'\chi_{23o}\right]-\frac{2\alpha\omega
d +\beta\bar{\mu}}{2\alpha
R_o+\beta\mu_o}\overset{~~\textrm{eff}}{{P}_{\phi o}},
\end{align}
where over bar shows perturbed form of Einstein tensors and are
written in Appendix as Eqs.(\ref{mbar})-(\ref{ppbar}). In case of
hydro-static equilibrium, the second dynamical equation has the
following form
\begin{align}\label{30}
&\frac{1}{A_o}\overset{~~\textrm{eff}}{{P_{ro}}'}+\frac{A_o'}{A_o^2}
\left(\overset{~~\textrm{eff}}{\mu_{o}}+\overset{~~\textrm{eff}}{P_{ro}}\right)
+\frac{B_o'}{A_oB_o}
\left(\overset{~~\textrm{eff}}{P_{ro}}-\overset{~~\textrm{eff}}{P_{zo}}\right)
+\frac{C_o'}{A_oC_o}
\left(\overset{~~\textrm{eff}}{P_{ro}}-\overset{~~\textrm{eff}}{P_{\phi
o}}\right)+Z_{2o}=0,
\end{align}
while, their non-static forms are
\begin{align}\label{31}
&\overset{~~\textrm{eff}}{\dot{\bar{\mu}}}+\dot{\omega}\eta=0,\\\nonumber
&\frac{1}{A_o}\left[\overset{~~\textrm{eff}}{\bar{P_{r}}'}+\frac{\omega
a}{A_o}\left\{\overset{~~\textrm{eff}}{P_{ro}'}+\frac{B_o'}{B_o}
\left(\overset{~~\textrm{eff}}{P_{zo}}-\overset{~~\textrm{eff}}{P_{ro}}\right)+\frac{C_o'}{C_o}
\left(\overset{~~\textrm{eff}}{P_{\phi
o}}-\overset{~~\textrm{eff}}{P_{ro}}\right)+
\left(\overset{~~\textrm{eff}}{\mu_{o}}+\overset{~~\textrm{eff}}{P_{ro}}\right)\right.\right.\\\nonumber
&\left.\left.\times
\left(\frac{a'}{a}-\frac{2A_o'}{A_o}\right)\right\}
+\omega\left\{\left(\overset{~~\textrm{eff}}{P_{ro}}-\overset{~~\textrm{eff}}{P_{zo}}\right)
\left(\frac{b}{B_o}\right)'+\left(\overset{~~\textrm{eff}}{P_{ro}}
-\overset{~~\textrm{eff}}{P_{\phi o}}\right)
\left(\frac{c}{C_o}\right)'-\frac{B_o'}{B_o}\right.\right.\\\label{32}
&\left.\left.\times\left(\overset{~~\textrm{eff}}{\bar{P_{z}}}-\overset{~~\textrm{eff}}{\bar{P_{r}}}\right)
-\left(\overset{~~\textrm{eff}}{\bar{P_{\phi}}}+\overset{~~\textrm{eff}}{\bar{P_{r}}}\right)
\frac{C_o'}{C_o}+\left(\overset{~~\textrm{eff}}{\bar{\mu}}+\overset{~~\textrm{eff}}{\bar{P_{r}}}\right)
\frac{A_o'}{A_o}\right\} \right]+\omega \bar{Z_2}=0,
\end{align}
where
\begin{align}\nonumber
&\eta=\overset{~~\textrm{eff}}{\mu_o}\left(\frac{a}{A_o}+\frac{b}{B_o}+\frac{c}{C_o}\right)
+\overset{~~\textrm{eff}}{P_{ro}}\left(\frac{a}{A_o}+\frac{b}{B_o}\right)
+\frac{c}{C_o}\overset{~~\textrm{eff}}{P_{\phi o}}+ A_o\bar{Z_1}.
\end{align}
Under non-static environment, the scalar variables associated with
expansion and shear tensors are found as follows
\begin{align}\nonumber
\bar{\Theta}&=\frac{\dot{\omega}}{A_o}\left(\frac{a}{A_o}+\frac{b}{B_o}+\frac{c}{C_o}\right),~~
\bar{\sigma_s}=\frac{\dot{\omega}}{A_o}\left(\frac{b}{B_o}-\frac{a}{A_o}\right)~~
\bar{\sigma_k}=\frac{\dot{\omega}}{A_o}\left(\frac{c}{C_o}-\frac{a}{A_o}\right).
\end{align}

\subsection{Stability Analysis}

Here, we want to discuss about the stability of cylindrical
anisotropic compact objects in terms of the stiffness parameter
$\Gamma_1$. The Harrison-Wheeler equation of state \cite{27} has a
great impact in this context which forms a relationship between
pressure components and energy density given as
\begin{equation}\label{33}
\bar{P_i}=\bar{\mu}\frac{P_{i0}}{\mu_0+P_{i0}}\Gamma_1.
\end{equation}
Then, Eq.(\ref{31}) can be rewritten as follows
\begin{align}\nonumber
\overset{~~\textrm{eff}}{\dot{\bar{\mu}}}=-\dot{\omega}\eta.
\end{align}
The integration of this equation gives
\begin{align}\label{34}
\overset{~~\textrm{eff}}{\bar{\mu}}=-\omega\eta.
\end{align}
Using the above value of $\bar{\mu}^{\textrm{eff}}$ in
Eq.(\ref{33}), we obtain
\begin{align}\label{35}
\overset{~~\textrm{eff}}{\bar{P}_{r}}=-\Gamma_1
\frac{\overset{~~\textrm{eff}}{P_{ro}}\eta\omega}{(\overset{~~\textrm{eff}}{\mu_{o}}+\overset{~~\textrm{eff}}{P_{ro}})},\quad
\overset{~~\textrm{eff}}{\bar{P}_{z}}=-\Gamma_1
\frac{\overset{~~\textrm{eff}}{P_{z
o}}\eta\omega}{(\overset{~~\textrm{eff}}{\mu_{o}}+\overset{~~\textrm{eff}}{P_{zo}})},\quad
\overset{~~\textrm{eff}}{\bar{P}_{\phi}}=-\Gamma_1
\frac{\overset{~~\textrm{eff}}{P_{\phi
o}}\eta\omega}{(\overset{~~\textrm{eff}}{\mu_{o}}+\overset{~~\textrm{eff}}{P_{\phi
o}})}.
\end{align}
Substituting the values from Eqs.(\ref{34}) and (\ref{35}) in
Eq.(\ref{30}), the corresponding modified collapse equation turns
out to be
\begin{align}\nonumber
&\Gamma_1\left[\frac{\overset{~~\textrm{eff}}{P_{ro}}}{\overset{~~\textrm{eff}}
{\mu_{o}}+\overset{~~\textrm{eff}}{P_{ro}}}
\left\{\eta\left(\frac{B_o'}{B_o}+\frac{C'_o}{C_o}-\frac{A_o'}{A_o}\right)-\eta'+\frac{\eta}
{(\overset{~~\textrm{eff}}{\mu_{o}}+\overset{~~\textrm{eff}}{P_{ro}})}
(\overset{~~\textrm{eff}}{\mu_{o}}'+\overset{~~\textrm{eff}}{P_{ro}}')\right\}
-\eta\right.\\\nonumber &\times\left.
\frac{\overset{~~\textrm{eff}}{P_{ro}}'}{\overset{~~\textrm{eff}}
{\mu_{o}}+\overset{~~\textrm{eff}}{P_{ro}}}+\eta\frac{B_o'}{B_o}
\frac{\overset{~~\textrm{eff}}{P_{zo}}}{\overset{~~\textrm{eff}}
{\mu_{o}}+\overset{~~\textrm{eff}}{P_{zo}}}+\eta\frac{C_o'}{C_o}
\frac{\overset{~~\textrm{eff}}{P_{\phi o}}}{\overset{~~\textrm{eff}}
{\mu_{o}}+\overset{~~\textrm{eff}}{P_{\phi
o}}}\right]=-\frac{a}{A_o}\overset{~~\textrm{eff}}{P_{ro}}'+\overset{~~\textrm{eff}}{P_{ro}}
\left\{\frac{aB_o'}{A_oB_o}\right.\\\nonumber
&\left.-\frac{C_o'}{C_o}+\frac{a'}{a}-\frac{2A'_o}{A_o}
-\left(\frac{b}{B_o}\right)'-\left(\frac{c}{C_o}\right)'\right\}+\overset{~~\textrm{eff}}{P_{zo}}
\left\{\left(\frac{b}{B_o}\right)'-\frac{aB_o'}{A_oB_o}\right\}
+\overset{~~\textrm{eff}}{P_{\phi o}}\\\label{36} &\times
\left\{\left(\frac{c}{C_o}\right)'-\frac{C_o'}{C_o}\right\}
+\overset{~~\textrm{eff}}{\mu_{o}}\left(\frac{a'}{a}-\frac{2A'_o}{A_o}\right)
+\frac{\eta A'_o}{A_o}+A_o\bar{Z_2}.
\end{align}
In a given equation, the terms including adiabatic index $\Gamma_1$
would generate pressure and counter gravitational effects while the
remaining terms work as the generator of the gravity force. The
effects, produced by principal stresses and $f(R,T,Q)$ gravity terms
intervened by fluid have greatest relevance in the analysis of
gravity forces.

\subsubsection{N Approximations}

Here, we compute the instability for cylindrical interior system at N
limit with the theory of gravity induced by $\alpha R^2+\beta Q$
model. In N regime, we shall consider a flat background metric, that provide weak field
approximations. Therefore, we take
\begin{equation*}
A_0=1, \quad B_0=1.
\end{equation*}
Since, we are dealing with the compact configurations of cosmic stellar filament, therefore, we assume that
the energy density of the matter content is much greater than the pressure components. Due to this reason, we shall
consider the following constraint in our calculation with N limit
\begin{equation*}
\mu_0\gg P_{i0}.
\end{equation*}
It was demonstrated by Chandrasekhar \cite{17} and Herrera \emph{et al.} \cite{18} that all the terms
coming in the stability conditions should be positive definite. Therefore, to attain the instability
regions of cylindrical stellar system, we are considering each term in the respective collapse equation to be
positive. The collapse equation (\ref{36}) takes the form
\begin{align}\nonumber
&\left[\overset{~~\textrm{eff}}{\mu_{o}}\left(a+b+\frac{c}{C_o}\right)+\bar{Z_1}\right]
\Gamma_1=\overset{~~\textrm{eff}}{\mu_{o}}(a'/a)+\Pi+\bar{Z_2},
\end{align}
where
\begin{align}\label{37}
\Pi=&b'\left(\overset{~~\textrm{eff}}{P_{zo}}-\overset{~~\textrm{eff}}
{P_{ro}}\right)+\left(\overset{~~\textrm{eff}}{P_{\phi
o}}+\overset{~~\textrm{eff}}{P_{ro}}\right)\frac{C'_o}{C_o}-\left(\frac{c}{C_o}\right)'
\left(\overset{~~\textrm{eff}}{P_{\phi
o}}+\overset{~~\textrm{eff}}{P_{ro}}\right)-a\overset{~~\textrm{eff}}{P_{ro}}',
\end{align}
includes anisotropic effects for onset of instability regimes in
cylindrical compact objects.

Now, we recall the work of Chandrasekhar \cite{17}, who checked the collapsing behavior of a perfect spherical star
with the help of numerical value of $\Gamma_1$. He found three possibilities about the N limits of the star. These are
\begin{enumerate}
  \item The effects of the star weight will be stronger than pressure, once the system satisfies
   $\Gamma_1<4/3$ condition. This would eventually lead the body to enter into collapse state.
  \item The initial compression would lead the system towards hydrostatic equilibrium, if $\Gamma_1=4/3$.
    \item Further, the limit $\Gamma_1>4/3$ indicates that the influence of pressure on the stellar dynamics is much greater than the star weight, thereby increasing the resulting outward force. Then, the system will move towards equilibrium and is said to be dynamical stable.
\end{enumerate}
Keeping in mind the same analysis for $f(R,T,Q)$ theory of gravity, the
evolving cylindrical anisotropic stellar object will be in phase of hydrostatic equilibrium
whenever it satisfies
\begin{align}\label{38}
&\Gamma_1=\frac{|\overset{~~\textrm{eff}}{\mu_{o}}(a'/a)+\Pi+\bar{Z_2}|}
{|\overset{~~\textrm{eff}}{\mu_{o}}\left(a+b+\frac{c}{C_o}\right)+\bar{Z_1}|}.
\end{align}
If the effects of $|\overset{~~\textrm{eff}}{\mu_{o}}(a'/a)+\Pi+\bar{Z_2}|$ and
$|\overset{~~\textrm{eff}}{\mu_{o}}\left(a+b+\frac{c}{C_o}\right)+\bar{Z_1}|$
are equal, then
\begin{align}\label{n1}
&\Gamma_1=1
\end{align}
will give us the condition of hydrostatic equilibrium for the cylindrically symmetric
anisotropic interiors. However, if the role of $|\overset{~~\textrm{eff}}{\mu_{o}}(a'/a)+\Pi+\bar{Z_2}|$ is lesser
than $|\overset{~~\textrm{eff}}{\mu_{o}}\left(a+b+\frac{c}{C_o}\right)+\bar{Z_1}|$,
then the relation
\begin{align}\label{39}
&\Gamma_1<\frac{|\overset{~~\textrm{eff}}{\mu_{o}}(a'/a)+\Pi+\bar{Z_2}|}
{|\overset{~~\textrm{eff}}{\mu_{o}}\left(a+b+\frac{c}{C_o}\right)+\bar{Z_1}|}
\end{align}
shows that the given system is in unstable region and the range of adiabatic index would belongs to
$(0,1)$. If the
modified gravity forces generated by
$|\overset{~~\textrm{eff}}{\mu_{o}}(a'/a)+\Pi+\bar{Z_2}|$ are higher
than that of
$|\overset{~~\textrm{eff}}{\mu_{o}}\left(a+b+\frac{c}{C_o}\right)+\bar{Z_1}|$, then
this will make the system enter into the stable window. This means that
the forces of anti-gravity and principal stresses produce the
stability constraint at N region as $$\Gamma_1>1.$$
This state is said to be the dynamical stable.

\subsubsection{pN Approximations}

In order to attain the pN instability constraints, we consider
$A_o(r)=1-\phi,~B_o(r)=1+\phi$ with effects upto $O(\phi)$, where
$\phi(r)=\frac{m_0}{r}$. In this context, the collapse equation (\ref{36})
provides the following value of $\Gamma_1$
\begin{align}\label{pn1}
\Gamma_1=\frac{F_{pN}}{E_{pN}},
\end{align}
where
\begin{align}\nonumber
F_{pN}&=\frac{\overset{~~\textrm{eff}}{P_{ro}}}{\overset{~~\textrm{eff}}
{\mu_{o}}+\overset{~~\textrm{eff}}{P_{ro}}}\left[-\eta'_{pN}+\eta_{pN}
\left\{\frac{C'_o}{C_o}+\frac{\overset{~~\textrm{eff}}{\mu_{o}'}}{\overset{~~\textrm{eff}}
{\mu_{o}}+\overset{~~\textrm{eff}}{P_{ro}}}+\phi'(1-\phi)\frac{\overset{~~\textrm{eff}}{P_{zo}}}
{\overset{~~\textrm{eff}}
{\mu_{o}}+\overset{~~\textrm{eff}}{P_{zo}}}+2\phi'\right.\right.\\\nonumber
&\left.\left.+\frac{C'_o}{C_o}
\frac{\overset{~~\textrm{eff}}{P_{\phi o}}}{\overset{~~\textrm{eff}}
{\mu_{o}}+\overset{~~\textrm{eff}}{P_{\phi
o}}}\right\}\right],\\\nonumber
E_{pN}&=-a(1+\phi)\overset{~~\textrm{eff}}{P_{ro}}'+S_1
(\overset{~~\textrm{eff}}{P_{ro}}-\overset{~~\textrm{eff}}{P_{zo}})
+S_2(\overset{~~\textrm{eff}}{P_{ro}}-\overset{~~\textrm{eff}}{P_{\phi
o}})+S_3(\overset{~~\textrm{eff}}{\mu_{o}}+\overset{~~\textrm{eff}}{P_{ro}}),\\\nonumber
&-\phi'(1+\phi)\eta_{pN}+(1-\phi)\bar{Z_2}.
\end{align}
The anisotropic cosmic filament will enter into the window of stable configurations, once the modified gravity
forces generated by $F_{pN}$ are greater than that of $E_{pN}$. In that case, the stability of the relativistic
system is ensured by the following pN limit
\begin{align}\label{pn2}
\Gamma_1>\frac{F_{pN}}{E_{pN}}.
\end{align}
However, if during evolution, the system attains the state at which $F_{pN}=E_{pN}$, then the system will cease
in the regime of equilibrium. At that time, the cylindrical system will no longer be in the evolutionary phases.
One can deal with such situation by considering Eq.(\ref{pn1}). The constraint for instability can be entertained by the
anisotropic cylindrical compact system, if the impact of $F_{pN}$ is less than $E_{pN}$. This would
give
\begin{align}\label{pn3}
\Gamma_1<\frac{F_{pN}}{E_{pN}}.
\end{align}
This pN instability limit depends upon the contribution of principal stresses
and counter gravity terms related with $\Gamma_1$ and $f(R,T,Q)$
gravity. This also indicates the significance of hydrostatic
equilibrium factors in the study of dynamical unstable regimes of
our system.

\section{Concluding Remarks}

In the framework of modified gravity, the stability problem of
massive objects has appeared as a main concern in relativistic
astrophysics. In this paper, we have analyzed the instability ranges
of self-gravitating cylindrical collapsing model in $f(R,T,Q)$
gravity structure. We have investigated the field equations for
cylindrical symmetric spacetime within anisotropic and
non-dissipative matter distribution. In this aspect, the dynamical
equations are developed by using the contraction of Bianchi
identities. The perturbed profile of the field, dynamical equations
and kinematical quantities are evaluated by imposing the
perturbation scheme on material and geometric variables.

Initially, we have supposed that our cylindrical system is in
hydrostatic equilibrium position. However, as time passes, it
undergoes into the oscillating phase. Therefore, the resulting
equations are applied to construct the collapse equation, which is
further analyzed at N and pN limits. In this background, adiabatic
index assisted by equation of state has been used to quantify the
stiffness of matter composition. Also, we have considered a feasible
model of $f(R,T,Q)$ theory and examined its impact in the dynamical
evolution of locally anisotropic celestial system. It is noticed
that additional curvature terms are appearing because of the
modification in the gravity model, which are the major cause of
obstacles in evolving celestial objects. Consequently, forming the
evolving cosmic filament system more stable due to their
non-attractive behavior.

It is noted that, for the stability of isotropic spherical
relativistic bodies, the particular numerical value of stiffness
parameter, i.e., $\frac{4}{3}$, was calculated by Chandrasekhar
\cite{17}. Since then, many astrophysicists have tried to examine
the instability regimes for various celestial geometries. We have
observed the critical role of adiabatic index in the description of
un-stable/stable regimes. We also examined that $\Gamma_1$ depends
upon the static configuration of geometry and matter as well as on
the additional terms which appear due to matter curvature coupling.
It is noted that the system will remain unstable whenever it holds
up its range as specified in expressions (\ref{39}) and (\ref{pn3})
for N and pN limits, respectively. When the system unable to follow
the above-mentioned ranges, it will enter into the stable or
equilibrium phase. It should be remarkably noted that in the absence
of non-minimal coupling of matter and geometry, these results mark
down to $f(R,T)$ outcomes. However, in case of vacuum, one can get
result of $f(R)$ gravity.

\vspace{0.25cm}

{\bf Acknowledgments}

\vspace{0.25cm}

The authors would like to thank the anonymous reviewer for valuable
and constructive comments and suggestions to improve the quality of
the paper.

\renewcommand{\theequation}{A\arabic{equation}}
\setcounter{equation}{0}
\section*{Appendix A}

The quantities $\chi_i$'s and $\psi$'s appearing in
Eqs.(\ref{11n})-(\ref{14n}) are
\begin{align}\nonumber
\chi_1&=1+f_T+f_Q\left[\frac{-3R}{2}+\frac{1}{A^2}\left\{\frac{4\dot{A}^2}{A^2}-2\frac{\ddot{A}}{A}+\frac{A'^2}{A^2}-\frac{A'^2}{A}
-\frac{\dot{A}^2}{A}+\frac{A''}{A}-\frac{\dot{A}}{A}\right.\right.\\\nonumber
&\left.\left.\times\left(\frac{\dot{B}}{B}+\frac{\dot{C}}{C}\right)+\frac{{A'}}{A}\left(\frac{{B'}}{B}+\frac{{C'}}{C}\right)\right\}\right]
+\frac{\dot{f_Q}}{A^2}\left(\frac{3\dot{A}}{2A}-\frac{\dot{B}}{2B}-\frac{\dot{C}}{2C}\right)+\frac{f_Q''}{2A^2}\\\nonumber
&+\frac{f_Q'}{2A^2}
\left(\frac{B'}{B}+\frac{C'}{C}\right),\\\nonumber
\chi_2&=\frac{f_Q}{A^2}\left(\frac{\dot{B}}{B}-\frac{9\dot{A}}{2A}-\frac{\dot{C}}{C}\right),\quad
\chi_3=\frac{1}{A^2}\left[f_Q''
+\frac{f_Q'}{2}\left(\frac{B'}{B}+\frac{C'}{C}\right)\right],\\\nonumber
\chi_4&=\frac{1}{A^2}\left[f''_Q+f_Q\left(\frac{4A'^2}{A^2}-\frac{A''}{A}-\frac{\dot{A}^2}{A^2}\right)-\frac{5A'}{A}f_Q'
-\frac{\dot{A}}{2A}\dot{f_Q}\right],\\\nonumber
\chi_5&=\frac{1}{A^2}\left[f_Q\left(\frac{\dot{B}^2}{B^2}-\frac{B'^2}{B^2}\right)-\frac{\dot{B}}{2B}\dot{f_Q}
+\frac{B'}{2B}f'_Q\right],\\\nonumber
\chi_6&=\frac{1}{A^2}\left[f_Q\left(\frac{\dot{C}^2}{C^2}-\frac{C'^2}{C^2}\right)-\frac{\dot{C}}{2C}\dot{f_Q}
+\frac{C'}{2C}f'_Q\right],\\\nonumber
\chi_7&=1+f_T-\frac{3R}{2}f_Q+\frac{f_Q}{A^2}\left[\frac{2A''}{A}-\frac{3A'^2}{A^2}-2\frac{\dot{A}^2}{A^2}
-\frac{\ddot{A}}{A}-\frac{\dot{A}}{A}\left(\frac{\dot{B}}{B}+\frac{\dot{C}}{C}\right)\right.\\\nonumber
&\left.
+\frac{A'}{A}\left(\frac{B'}{B}+\frac{C'}{C}\right)\right]+\frac{f_Q'}{A^2}\left(\frac{7A'}{A}
+\frac{B'}{2B}+\frac{C'}{2C}\right)-\frac{\dot{f_Q}}{2A^2}\left(\frac{3\dot{A}}{A}
+\frac{2\dot{B}}{B}\right.\\\nonumber
&\left.+\frac{\dot{C}}{2C}\right)-\frac{\ddot{f_Q}}{2A^2},\quad
\chi_{11}=\frac{-f_Q}{2A^2}\left(3\frac{\dot{A}}{A}+\frac{\dot{B}}{B}+\frac{\dot{C}}{C}\right)-\frac{\dot{f_Q}}{A^2}\\\nonumber
\chi_9&=\frac{f_Q}{A^2}\left(\frac{\ddot{A}}{A}-4\frac{\dot{A}^2}{A^2}-\frac{A'^2}{A^2}\right)
+\frac{1}{2A^2}\left(\frac{5\dot{A}}{A}\dot{f_Q}-\ddot{f_Q}+\frac{A'}{A}f_Q'\right),\\\nonumber
\chi_{10}&=\frac{f_Q}{A^2}\left(\frac{{C'^2}}{C^2}-\frac{\dot{C}^2}{C^2}\right)
+\frac{1}{2A^2}\left(\frac{\dot{A}}{A}\dot{f_Q}-\frac{C'}{C}f_Q'\right),\\\nonumber
\chi_{12}&=\frac{f_Q}{2A^2}\left(\frac{9{A'}}{A}+\frac{{B'}}{B}+\frac{{C'}}{C}\right)-\frac{2{f_Q'}}{A^2},\\\nonumber
\chi_{13}&=\frac{f_Q}{A^2}\left(\frac{\ddot{A}}{A}-\frac{4\dot{A}^2}{A^2}-\frac{A'^2}{A^2}\right)+\frac{1}{2A^2}
\left(\frac{5\dot{A}}{A}\dot{f_Q}+\frac{A'}{A}f_Q'-\ddot{f_Q}\right),\\\nonumber
\chi_{14}&=\frac{1}{2A^2}\left(\frac{5\dot{A}}{A}f_Q-2\dot{f_Q}\right),\\\nonumber
\chi_{15}&=\frac{f_Q}{A^2}\left(\frac{A''}{A}-\frac{4A'^2}{A^2}-\frac{\dot{A}^2}{A^2}\right)
+\frac{1}{A^2}\left(\frac{5A'}{A}f'_Q+\frac{\dot{A}}{2A}\dot{f_Q}-\frac{f''_Q}{2}\right),\\\nonumber
\chi_{16}&=\frac{f_Q}{A^2}\left\{\frac{B''}{B}-\frac{\ddot{B}}{2B}-\frac{\dot{B}}{B}\left(\frac{\dot{C}}{C}+\frac{4\dot{B}}{B}\right)
+\frac{B'}{B}\left(\frac{C'}{C}+\frac{3B'}{B}\right)-\frac{\dot{f_Q}}{A^2}\left(\frac{4\dot{B}}{B}\right.\right.\\\nonumber
&\left.\left.+\frac{\dot{C}}{2C}\right)\right\}+1+f_T-\frac{3Rf_Q}{2}+\frac{1}{2A^2}(f_Q''-\ddot{f_Q}),\\\nonumber
\chi_{17}&=\frac{f_Q}{A^2}\left(\frac{4B'}{B}+\frac{C'}{C}\right)+\frac{f'_Q}{A^2},~\chi_{18}=\frac{f_Q}{A^2}
\left(\frac{2\dot{B}}{B}+\frac{\dot{C}}{2C}\right)+\frac{\dot{f_Q}}{A^2}\\\nonumber
\chi_{19}&=\frac{f_Q}{A^2}
\left(\frac{C'^2}{C^2}-\frac{\dot{C}^2}{C^2}\right)+\frac{1}{2A^2}\left(\frac{\dot{C}}{C}\dot{f_Q}-\frac{C'}{C}f'_Q\right),\\\nonumber
\chi_{20}&=\frac{f_Q}{A^2}\left(\frac{B'^2}{B^2}-\frac{\dot{B}^2}{B^2}\right)+\frac{1}{2A^2}\left(\frac{\dot{B}}{B}\dot{f_Q}
-\frac{B'}{B}f_Q'\right),\\\nonumber
\chi_{21}&=\frac{f_Q}{A^2}\left\{\frac{C''}{C}+\frac{C'}{C}\left(\frac{B'}{B}+\frac{C'}{C}\right)
-\frac{\dot{C}}{C}\left(\frac{\dot{B}}{B}+4\frac{\dot{C}}{C}\right)-\frac{\ddot{C}}{C}\right\}
+\frac{f_Q'}{2A^2} \left(\frac{4C'}{C}\right.\\\nonumber
&\left.+\frac{B'}{B}\right)-\frac{\dot{f_Q}}{2A^2}\left(\frac{\dot{B}}{B}+\frac{5\dot{C}}{C}\right)
+\frac{1}{2A^2}(f_Q''-\ddot{f_Q})+1+f_T-\frac{3R}{2}f_Q,\\\nonumber
\chi_{22}&=-\frac{\dot{f_Q}}{A^2}-\frac{f_Q}{2A^2}\left(\frac{\dot{B}}{B}+\frac{5\dot{C}}{C}\right),~
\chi_{23}=-\frac{f_Q'}{A^2}+\frac{f_Q}{2A^2}\left(\frac{B'}{B}+\frac{5C'}{C}\right),\\\nonumber
\psi_1&=\frac{1}{A^2}\left(\frac{\dot{A}}{A}+\frac{\dot{B}}{B}+\frac{\dot{C}}{C}\right),\quad
\psi_2=\frac{1}{A^2}\left(\frac{{A'}}{A}-\frac{{B'}}{B}-\frac{{C'}}{C}\right),\\\nonumber
\psi_3&=\frac{1}{A^2}\left(\ddot{f_R}-f_R''+\frac{\dot{C}}{C}\dot{f_R}+\frac{C'}{C}f'_R\right),~
\psi_4=\frac{1}{A^2}\left(\ddot{f_R}-f_R''+\frac{\dot{B}}{B}\dot{f_R}+\frac{B'}{B}f'_R\right).
\end{align}
The expressions $S_i$'s appearing in Eq.(\ref{pn1}) are
\begin{align}\nonumber
S_1&=a\phi'-[b(1-\phi)]',~S_2=\frac{C'_o}{C_o}-\left(\frac{c}{C_o}\right)',~S_3=\frac{a'}{a}+2\phi'(1+\phi).
\end{align}
The quantities $Z_1$ and $Z_2$ coming in Eqs.(\ref{17}) and
(\ref{18}) are
\begin{align}\nonumber
Z_1&=\frac{2}{1+Rf_{RT}+2f_T}\left[2(\mu
f_T\dot{)}-(f_{RT}R^{00}\mu\dot{)}+(P_rf_{RT}R^{10})'-\left\{\frac{1}{A^2}(\mu+P_r)\right.\right.\\\nonumber
&\left.\left.+\frac{P_z}{B^2}
+\frac{P_\phi}{C^2}\right\}_{,0}\{f_{RT}\sum_{i=0}^3R_{ii}+f_T(B^2+C^2)\}-G^{00}(\mu
f_T\dot{)}-\frac{\mu}{2A^2}(Rf_{RT}\dot{)}-\frac{\mu\dot{f_T}}{A^2}\right.\\\label{z1}
&\left.-G^{10}(\mu f_T)'\right],\\\nonumber
Z_2&=\frac{2}{1+Rf_{RT}+2f_T}\left[2(\mu
f_T)'-(f_{RT}R^{01}\mu\dot{)}+(P_rf_{RT}R^{11})'-\{\frac{1}{A^2}(\mu+P_r)\right.\\\nonumber
&\left.+\frac{P_z}{B^2}
+\frac{P_\phi}{C^2}\}'\{f_{RT}\sum_{i=0}^3R_{ii}+f_T(B^2+C^2)\}-G^{01}(\mu
f_T\dot{)}-G^{11}(\mu f_T)'-\frac{P_r}{2A^2}\right.\\\label{z2}
&\left.\times\{(Rf_{RT})'+2f_T'\}\right].
\end{align}
The static configurations of Einstein tensors appearing in
Eqs.(\ref{23})-(\ref{25}) are
\begin{align}\label{sm}
G_{00}^{(S)}&=\frac{1}{A_o^2}\left[\frac{A_o'}{A_o}\left(\frac{B_o'}{B_o}+\frac{C_o'}{B_o}\right)-\frac{B_o''}{B_o}-\frac{C_o''}{C_o}
-\frac{B_o'C_o'}{B_oC_o}\right],\\\label{sp}
G_{11}^{(S)}&=\frac{1}{A_o^2}\left[\frac{A_o'}{A_o}\left(\frac{B_o'}{B_o}+\frac{C_o'}{B_o}\right)+
\frac{B_o'C_o'}{B_oC_o} \right],\\\label{spp}
G_{22}^{(S)}&=\frac{1}{A_o^2}\left[\frac{A_o''}{A_o}+\frac{C_o''}{C_o}-\frac{A_o'^2}{A_o^2}\right],\quad
G_{33}^{(S)}=\frac{1}{A_o^2}\left[\frac{A_o''}{A_o}+\frac{B_o''}{B_o}-\frac{A_o'^2}{A_o^2}\right].
\end{align}
The non-static perturbed configurations of Einstein tensors
appearing in Eqs.(\ref{26})-(\ref{29}) are
\begin{align}\nonumber
\bar{G_{00}}&=\frac{\omega}{A_o^2}\left[\frac{bB_o''}{B_o}-\frac{b''}{B_o}+\frac{cC_o''}{C_o}
+\left(\frac{a}{A_o}\right)'\left(\frac{B_o'}{B_o}+\frac{C_o'}{C_o}\right)+\frac{A_o'}{A_o}
\left\{\left(\frac{b}{B_o}\right)'\right.\right.\\\label{mbar}
&\left.\left.+\left(\frac{c}{C_o}\right)'\right\}-\frac{C_o'}{C_o}
\left(\frac{b}{B_o}\right)'-\frac{B_o'}{B_o}\left(\frac{c}{C_o}\right)'\right]-2\overset{~~\textrm{eff}}{\mu_o}\omega
\frac{a}{A_o},\\\nonumber \bar{G_{11}}&=-\frac{\ddot{\omega}}{A_o^2}
\left(\frac{b}{B_o}+\frac{c}{C_o}\right)+\frac{\omega}{A_o^2}\left[
\left(\frac{B_o'}{B_o}+\frac{C_o'}{C_o}\right)\left(\frac{a}{A_o}\right)'+
\left(\frac{A_o'}{A_o}+\frac{C_o'}{C_o}\right)\left(\frac{b}{B_o}\right)'\right.\\\label{prbar}
&\left.+\left(\frac{B_o'}{B_o}+\frac{A_o'}{A_o}\right)\left(\frac{c}{C_o}\right)'\right]
-2\overset{~~\textrm{eff}}{P_{ro}}\frac{a}{A_o}\omega,\\\nonumber
\bar{G_{22}}&=-\frac{\ddot{\omega}}{A_o^2}
\left(\frac{a}{A_o}+\frac{c}{C_o}\right)+\frac{\omega}{A_o^2}\left[
\frac{a''}{A_o}+\frac{c''}{C_o}-\frac{aA_0''}{A_o^2}-\frac{cC_o''}{C_o^2}
-\frac{2A_o'}{A_o}\left(\frac{a}{A_o}\right)'\right]\\\label{pzbar}
&-2\overset{~~\textrm{eff}}{P_{zo}}\frac{a}{A_o}\omega,\\\nonumber
\bar{G_{33}}&=-\frac{\ddot{\omega}}{A_o^2}
\left(\frac{a}{A_o}+\frac{b}{B_o}\right)+\frac{\omega}{A_o^2}\left[
\frac{a''}{A_o}+\frac{b''}{B_o}-\frac{aA_0''}{A_o^2}-\frac{bB_o''}{B_o^2}
-\frac{2A_o'}{A_o}\left(\frac{a}{A_o}\right)'\right]\\\label{ppbar}
&-2\overset{~~\textrm{eff}}{P_{\phi o}}\frac{a}{A_o}\omega.
\end{align}


\vspace{0.25cm}

\begin{thebibliography}{40}

\bibitem{snIa} A. G. Riess et al. (Supernova Search Team Collaboration), Astron. J. 116,
1009 (1998).

\bibitem{snIb} S. Perlmutter et al. (Supernova Cosmology Project Collaboration),
Astrophys. J. 517, 565 (1999).

\bibitem{cmb1} R. R. Caldwell and M. Doran, Phys. Rev. D \textbf{69}, 103517 (2004).

\bibitem{cmb2} T. Koivisto and D. F. Mota, Phys. Rev. D \textbf{73}, 083502 (2006).

\bibitem{z5a} S. Nojiri and S. D. Odintsov, Phys. Lett. B, \textbf{599}, 137
(2004).

\bibitem{z5b} G. Cognola, E. Elizalde, S. Nojiri, S. D. Odintsov, L. Sebastiani and
S. Zerbini, Phys. Rev. D \textbf{77}, 046009 (2008).

\bibitem{b2b} S. Nojiri and S. D. Odintsov, Phys. Rep. \textbf{505}, 59 (2011).

\bibitem{z5d} K.~Bamba, S.~Capozziello, S.~Nojiri and S.~D.~Odintsov, Astrophys.\
Space Sci.\  {\bf 342}, 155 (2012).

\bibitem{6} R. Durrer and R. Maartens, arXiv:0811.4132 [astro-ph].

\bibitem{7} M.~Z.~Bhatti,~Z.~Yousaf~and~S.~Hanif,~Mod.~Phys.~Lett.~A~\textbf{32},~1750042~(2017).

\bibitem{8} T. Harko, F. S. N. Lobo, S. Nojiri, and S. D. Odintsov, Phys. Rev. D \textbf{84}, 024020 (2011).

\bibitem{8a} O. Bertolami and M. C. Sequeira, Phys. Rev. D \textbf{79}, 104010
(2009).

\bibitem{19} Z. Haghani, T. Harko, F. S. N. Lobo, H. R. Sepangi and S. Shahidi, Phys. Rev. D \textbf{88}, 044023 (2013).

\bibitem{20} S. D. Odintsov and D. S\'{a}ez-G\'{o}mez, Phys. Lett. B \textbf{725}, 437 (2013).

\bibitem{z20a} E. Elizalde and S. I. Vacaru, Gen. Relativ. Gravit. \textbf{47}, 64
(2015).

\bibitem{z20b} E. H. Baffou, M. J. S. Houndjo and J. Tosssa, Astrophys. Space Sci.
\textbf{361}, 376 (2016).

\bibitem{17} S. Chandrasekhar, Astrophys. J. \textbf{ 140}, 417 (1964).

\bibitem{18} L. Herrera, G. Le Denmat and N. O. Santos, Gen. Relativ. Gravit. \textbf{44}, 1143 (2012).

\bibitem{14} J. A. R. Cembranos, \'{A}. de la Cruz-Dombriz and B. M. N\'{u}\~{n}ez,
J. Cosmol. Astropart. Phys. \textbf{04}, 021 (2012).

\bibitem{10} Z. Yousaf, K. Bamba and M. Z. Bhatti, Phys. Rev. D \textbf{93}, 064059 (2016) [arXiv1603.03175 [gr-qc]];
Phys. Rev. D \textbf{93}, 124048 (2016) [arXiv1606.00147 [gr-qc]].

\bibitem{zs1} Z. Yousaf, Eur. Phys. J. Plus \textbf{132}, 71 (2017).

\bibitem{bin4} S. Chandrasekhar and E. Fermi, Astrophys. J. \textbf{118},
116 (1953).

\bibitem{bin6} J. Ostriker, Astrophys. J. \textbf{140}, 1529 (1964).

\bibitem{bin7} A. M. Fridman and V. L. Polyachenko, \emph{Physics of
Gravitating Systems}, (New York: Springer-Verlag, 1984).

\bibitem{bin5} B. G. Elmegreen, Astrophys. J. \textbf{231}, 372
(1979); ibid. 1994, Astrophys. J. \textbf{433}, 39 (1994).

\bibitem{bin3} J. Comparetta and A. C. Quillen, Mon. Not. R. Astron. Soc.
\textbf{414}, 810 (2011).

\bibitem{bin7a} N. Moeckel and A. Burkert, Astrophys. J. \textbf{807}, 67
(2015).

\bibitem{bin1} J. Binney and S. Tremaine, \emph{Galactic Dynamics, Princeton
Series in Astrophysics} (Princeton University Press, 1987).

\bibitem{bin2} P. H. Chavanis, Astron. Astrophys. \textbf{451}, 109 (2006).

\bibitem{bin8} A. C. Quillen and J. Comparetta, arXiv:1002.4870
[astro-ph.CO].

\bibitem{bin10} P. C. Myers, Astrophys. J. \textbf{764}, 140 (2013).

\bibitem{bin11} P. C. Breysse, M. Kamionkowski and A. Benson, Mon. Not. R. Astron. Soc.
\textbf{437}, 2675 (2014).

\bibitem{chi4} M. Sharif and R. Manzoor, Eur. Phys. J. C \textbf{76}, 330 (2016).

\bibitem{bin9} Y. Birnboim, D. Padnos and E. Zinger, Astrophys. J. Lett.
\textbf{832}, L4 (2016).

\bibitem{22} Z. Yousaf, M. Z. Bhatti and U. Farwa, Mon. Not. R. Astron. Soc. \textbf{464}, 4509 (2017).

\bibitem{v34} A. Einstein and N. Rosen, J. Franklin Inst. \textbf{223}, 43,
(1937).

\bibitem{sano1} J. M. M. Senovilla, Phys. Rev. D \textbf{88}, 064015
(2013).

\bibitem{23} I. Ayuso, J. B. Jim\'{e}nez, and \'{A}. de la Cruz-Dombriz, Phys. Rev. D \textbf{91}, 104003 (2015).

\bibitem{24} J. A. R. Cembranos, Phys. Rev. Lett. \textbf{102}, 141301 (2009).

\bibitem{v36} G. Darmois, \textit{Memorial des Sciences Mathematiques} (Gautheir-Villars,
Paris, 1927), Fasc. 25.

\bibitem{clif1} T.~Clifton, P.~Dunsby, R.~Goswami and A.~M.~Nzioki, Phys.\ Rev.\ D {\bf 87}, no. 6, 063517 (2013)[arXiv:1210.0730 [gr-qc]].

\bibitem{clif2} N. Deruelle, M. Sasaki and Y. Sendouda, Prog. Theor. Phys. \textbf{119}, 237 (2008).

\bibitem{chi1} W. B. Bonnor and N. O. Santos, Phys. Rep.
\textbf{181}, 269 (1989).

\bibitem{chi1a} L. Herrera, N. O. Santos, G. Le Denmat, Mon. Not. R. Astron. Soc. \textbf{237},
257 (1989).

\bibitem{chi2} R. Chan, L. Herrera and N. O. Santos, Mon. Not. R. Astron. Soc.
\textbf{265}, 533 (1993).

\bibitem{chi3} R. Chan, Mon. Not. R. Astron. Soc. \textbf{316}, 588 (2000).

\bibitem{chi4a} R. Chan, L. Herrera and N. O. Santos, Mon. Not. R. Astron. Soc.
\textbf{267}, 637 (1994).

\bibitem{chi5} H. R. Kausar and I. Noureen, Eur. Phys. J. C \textbf{74}, 2760 (2014).

\bibitem{27} B. K. Harrison, K. S. Throne, M. Wakano and J. A. Wheeler,
{\it Gravitation Theory and Gravitational Collapse} (University of Chicago press, 1965).

\end{thebibliography}
\end{document}